\documentclass{article}
\usepackage[utf8]{inputenc}
\usepackage{authblk}
\usepackage{float}
\usepackage{xcolor}
\usepackage{tikz}
\usepackage{soul}
\usepackage{multicol}
\usepackage{smartdiagram}
\usepackage{caption}
\usepackage{arev}
\usetikzlibrary{mindmap}
\usepackage{csvsimple}
\usepackage{pdflscape}
\usepackage{hyperref}
\usepackage{subcaption}
\usepackage{graphicx}
\usepackage[most]{tcolorbox}
\usepackage{rotating}
\usepackage{comment}
\usepackage{multirow}
\usepackage{lineno}
\usepackage{array}
\usepackage{xcolor}
\usepackage{lscape}
\usepackage[backend=biber,style=vancouver,sorting=none]{biblatex}
\usepackage{setspace}
\DeclareUnicodeCharacter{0301}{\'{o}}
\usepackage{amssymb}
\usepackage[a4paper, margin=1in]{geometry}
\usepackage{times}

\makeatletter
\renewcommand{\maketitle}{\bgroup\setlength{\parindent}{0pt}
\begin{flushleft}
  \textbf{\@title}

  \@author
\end{flushleft}\egroup
}
\makeatother

\providecommand{\keywords}[1]
{
  \textbf{\textit{Keywords---}} #1
}

\font\myfont=cmr12 at 17pt

\title{\myfont{Regulatory Approval Is Not Enough: Gaps in Trustworthy AI Reporting in FDA-Cleared Medical Devices}}

\date{}
\author[1,2]{Ahmed M Salih}
\author[3,4]{Oliver Díaz}
\author[3]{Alejandro Guzman}
\author[3,4]{Noah Marquez Vara}
\author[3]{Fotios Avgoustidis}
\author[3]{Rituraj Singh}
\author[5]{Saman Barakat}
\author[6]{Zahra Raisi-Estabragh}
\author[3,7]{Karim Lekadir}
\affil[1]{Division of Cardiovascular Sciences, University of Leicester, Leicester LE1 7RH, UK}
\affil[2]{British Heart Foundation (BHF) Leicester Centre of Research Excellence, University of Leicester, UK}
\affil[3]{Department de Matemàtiques i Informàtica, Universitat de Barcelona, Barcelona, Spain}
\affil[4]{Department of Computer Science and Engineering, Chalmers University of Technology, Gothenburg, Sweden}
\affil[5]{SCORE Lab, I3US Institute, Universidad de Sevilla, Seville, Spain}
\affil[6]{William Harvey Research Institute, NIHR Barts Biomedical Research Centre, Queen Mary University of London, Charterhouse Square, London, EC1M 6BQ, London, UK}
\affil[7]{Institució Catalana de Recerca i Estudis Avançats (ICREA), Barcelona, Spain}

\begin{document}
\maketitle
\thispagestyle{empty} 

\noindent
\begin{abstract}
\noindent
\textbf{Background:} 
Artificial intelligence and machine learning (AI/ML)-enabled medical devices are increasingly deployed in healthcare under evolving regulatory frameworks. As these systems become more integrated into clinical decision-making, there is growing expectation that they demonstrate key dimensions of trustworthy AI to support clinician, patient, and public trust. However, whether publicly available regulatory documentation provides sufficient evidence to independently assess the trustworthiness of cleared AI systems remains unclear.\\
\textbf{Methods:} 
We analysed FDA AI/ML-enabled medical device summary reports published between 2021 and 2025. Reports underwent automated keyword screening followed by multi-stage manual consensus review to identify documented evidence related to the six FUTURE-AI principles: Fairness, Universality, Traceability, Usability, Robustness, and Explainability. Descriptive, temporal, and clinical-domain analyses were performed. Multivariable logistic regression assessed whether year of clearance or clinical domain predicted higher reporting transparency, defined as evidence reported for three or more principles.\\
\textbf{Results:} 
Of 1,105 FDA summary reports screened, 519 were included. Trustworthy AI reporting was limited and uneven. Nearly one quarter of reports (24·7\%) provided no evidence for any principle, and no report documented evidence across all six. Robustness was most frequently reported (57·6\%), while Traceability (8·3\%) and Explainability (3·5\%) represented the most pronounced gaps. Neither year of clearance (OR 1·02, 95\% CI 0·88–1·19; p=0·71) nor clinical domain (OR 0·73, 95\% CI 0·46–1·15; p=0·17) were significantly associated with higher reporting transparency.\\
\textbf{Interpretation:} 
Substantial and persistent trustworthy AI reporting gaps exist in FDA regulatory documentation, revealing a disconnect between emerging trustworthy AI expectations and publicly disclosed evidence. Regulatory approval alone should not be considered a proxy for trustworthiness. Standardised, audit-ready reporting across the AI lifecycle is needed to support independent assessment and responsible adoption of healthcare AI.\\
\textbf{Funding:}
European Union’s Horizon Europe research and innovation programme under grant agreement No 101233553 (COMPASS-AI); Leicester City Football Club, from British Heart Foundation (RE/24/130031), Ministry of Science and Innovation of Spain (PID2023-146786OB-I00); European Union’s Horizon Europe (Marie Skłodowska-Curie grant agreement No 101211445 (BRIDGE project)).
\end{abstract}
\section{Introduction}
Trustworthy AI has become an important concept in the development and deployment of artificial intelligence and machine learning (AI/ML)-enabled medical devices in healthcare. As these technologies increasingly support clinical decision-making, they are expected to meet not only technical performance requirements but also broader ethical and societal expectations such as safety, fairness, explainability, accountability, and traceability. In the United States, the U.S. Food and Drug Administration (FDA) is the primary authority responsible for evaluating and clearing AI/ML-enabled medical devices through pathways such as 510(k), De Novo, and Premarket Approval (PMA)~\cite{fdaUSFood}. As part of this process, manufacturers submit summary reports describing the device’s intended use, clinical application, and validation evidence. These reports are publicly available through the FDA database~\cite{Center_for_Devices2026} and serve as an important transparency mechanism, allowing healthcare providers and clinicians to review the evidence supporting approved AI systems.\\
Regulatory and policy initiatives have increasingly incorporated principles related to trustworthy AI in healthcare. In 2019, the FDA proposed a regulatory framework for AI/ML-based Software as a Medical Device (SaMD), emphasizing lifecycle oversight and transparency in algorithm modifications. This was followed in 2021 by the introduction of the Good Machine Learning Practice (GMLP) guiding principles by the FDA, Health Canada, and the United Kingdom’s Medicines and Healthcare products Regulatory Agency (MHRA), highlighting key considerations such as data quality, robust validation, fairness, interpretability, and human-centred design ~\cite{FDA2}. Broader international governance efforts have also emerged. In the United States, Executive Order 14110 established a federal framework for the safe, secure, and trustworthy development of AI ~\cite{FDA3}. In parallel, the European Union introduced the EU AI Act, which defines regulatory obligations for high-risk AI systems, including medical devices, with requirements for transparency, accountability, and human oversight ~\cite{EU_AI_Act}. Within this evolving landscape, the FUTURE-AI international consortium proposed a comprehensive consensus guideline for trustworthy AI in healthcare, structured around six core principles: \textit{Fairness}, \textit{Universality}, \textit{Traceability}, \textit{Usability}, \textit{Robustness}, and \textit{Explainability}~\cite{lekadir2025future}.\\ 
Despite growing regulatory and societal emphasis on trustworthy AI, it remains unclear how these expectations are reflected in publicly available FDA documentation for AI/ML-enabled medical devices. A few studies have already shown gaps in specific aspects of transparency, such as demographic reporting, regulatory characteristics, or dataset disclosure ~\cite{mehta2025evaluating, almarie2025machine}. For example, a recent review of 692 FDA-cleared devices found that racial or ethnic composition was reported in only 3.6\% of validation datasets, while age information was absent in 81.6\% of reports and socioeconomic characteristics in 99.1\% \cite{muralidharan2024scoping}. Although these studies highlight important transparency concerns, they do not assess the broader spectrum of trustworthy AI principles such as explainability, traceability, usability, robustness, fairness, and generalisability. Consequently, the extent to which current FDA public summaries provide meaningful evidence aligned with emerging frameworks such as GMLP and FUTURE-AI remains largely unknown.\\
In this study, we establish the first systematic benchmark of trustworthy AI evidence reporting across publicly available FDA summaries of AI/ML-enabled medical devices. Using the FUTURE-AI framework as a structured, multi-dimensional lens, we assess the extent to which current regulatory documentation supports transparent evaluation of key trustworthy AI dimensions. Our analysis therefore shifts attention from device approval alone to the quality and completeness of the evidence communicated to clinicians, researchers, and healthcare systems.
\section{Methods}
\subsection{Data Sources and Eligibility}
Our analysis focused on devices cleared from 2021 onward, corresponding to the 5-year period following the introduction of the FDA AI/ML Action Plan and the joint GMLP ~\cite{FDA_regu}. This timeframe ensures the assessment of the most recent AI/ML-enabled medical devices developed under increasing regulatory and scientific awareness of trustworthy AI concepts. Custom Python scripts (Python 3.12.7; PyPDF2 version 3.0.1) were implemented to automatically download FDA-cleared AI/ML-enabled medical devices summaries in PDF format and systematically screen the reports for trustworthy AI-related keywords derived from the FUTURE-AI framework, including \textit{explainable, explainability, interpretable, interpretability, trustworthy, fairness, robustness, universality, generalisability, usability, accountability, transparency, adaptability, interoperability, understandability, bias, reproducibility, repeatability, and traceability.} The automated parsing identified the exact page number of each keyword occurrence, providing the basis for subsequent manual review and qualitative assessment.
\subsection{Systematic Review and Data Extraction}
The extraction process was conducted by a team of seven researchers with expertise in trustworthy AI principles. To ensure consistency and methodological rigor, the team adhered to the following multi-stage consensus protocol:

\begin{itemize}
    \item \textbf{Framework Development:} The research team initially reviewed a subset of the included reports to capture the diversity of terminology and reporting styles used by different manufacturers. Based on this preliminary assessment, the team collectively developed and refined the categorization framework detailed in Supplementary Table~\ref{TAI_ev}. This process ensured that the FUTURE-AI principles were consistently mapped to the specific terminology and regulatory language used in 510(k), De Novo, and PMA summary reports.
    
    \item \textbf{Harmonization and Consistency:} Throughout the manual review process, the research team held regular weekly meetings to discuss ambiguities encountered in the FDA summaries and ensure consistent interpretation of trustworthy AI concepts across reviewers. These discussions focused on distinguishing between standard regulatory compliance and meaningful trustworthy AI evidence. For example, the team differentiated general usability compliance references (e.g. IEC 62366) from AI-specific usability evidence, and clarified the identification of explicit explainability methods such as SHapley Additive exPlanations (SHAP). This iterative harmonization process ensured consistent application of the evaluation criteria across all reviewed reports.    
   
    \item \textbf{Principle-Specific Evidence Evaluation:} During the manual review stage, each keyword identified through automated screening was manually evaluated to determine whether it represented meaningful evidence rather than a superficial mention. To ensure consistency and reproducibility across the dataset, all reviewers assessed the context and evidence for each concept and assigned a positive or negative score according to the consensus FUTURE-AI criteria summarized in Table~\ref{tab:future_ai_criteria}.\\ 
   
\end{itemize}

\begin{table}[ht]
\centering
\caption{Principle-specific evaluation criteria used during manual review.}
\label{tab:future_ai_criteria}
\begin{tabular}{p{3cm} p{10cm}}
\hline
\textbf{Principle} & \textbf{Evidence required for a positive score} \\
\hline

\textbf{Fairness} & Subgroup analyses, bias mitigation strategies, or evaluation of performance disparities across populations \\

\textbf{Universality} & Validation across multiple cohorts, institutions, medical devices, or external datasets \\

\textbf{Traceability} & Auditability mechanisms, including data and model provenance, quality control, continuous evaluation, or logging procedure \\

\textbf{Usability} & Explicit consideration of human-AI interactions, clinical workflow integration, clinician feedback, or usability testing \\

\textbf{Robustness} & Stress testing, uncertainty estimation, missing data handling, robustness to noise, or evaluation across heterogeneous acquisitions \\

\textbf{Explainability} & Interpretable model outputs or feature-level insights, such as saliency maps, attention mechanisms, or highlighted regions of importance \\

\hline
\end{tabular}
\end{table}
\noindent
To validate the keyword-based search strategy, a manual review was additionally performed on reports that did not initially trigger any keywords. This cross-check ensured that relevant trustworthy AI-related evidence was not overlooked due to heterogeneous or non-standardized manufacturer terminology.\\
It is important to note that this study evaluates the transparency and completeness of publicly available FDA summary reports rather than the intrinsic performance or compliance of the AI/ML devices themselves. Therefore, the lack of evidence for a given trustworthy AI principle may reflect either a lack of compliance with that principle or insufficient public reporting of the corresponding evidence, both of which may negatively impact public trust.
\subsection{Data Analysis}
Various analyses were performed to assess in detail the reporting of trustworthy AI evidence across FDA AI/ML-enabled medical device summary reports and to identify patterns associated with reporting transparency. First, descriptive analyses were conducted to quantify the frequency of evidence reported for each FUTURE-AI principle across the full dataset, allowing identification of the most and least commonly addressed trustworthy AI dimensions. Temporal trend analyses were then performed according to year of FDA clearance to assess whether reporting practices have improved over time alongside the increasing maturity of AI regulatory frameworks and growing awareness of trustworthy AI principles. Comparisons across clinical domains, particularly between radiology and non-radiology devices, were additionally conducted to evaluate whether reporting transparency differed across specialties. To further investigate reporting patterns, co-occurrence analyses were performed to assess how frequently multiple FUTURE-AI principles were reported within the same summary report. \\
Finally, to identify factors associated with higher-quality reporting, we defined a binary metric termed \textit{High Reporting Transparency}, corresponding to reports containing evidence for three or more FUTURE-AI principles. Multivariable logistic regression was subsequently performed to evaluate associations between this metric and (1) year of FDA clearance (continuous variable) and (2) clinical domain (radiology versus non-radiology). Odds ratios (OR) with 95\% confidence intervals (CI) were calculated. All statistical analyses were performed using Python (version 3.12.7) with the \texttt{statsmodels} library. A two-sided p-value of $<0.05$ was considered statistically significant.

\section{Results}
\begin{figure}[H]
    \centering
    \includegraphics[width=\linewidth, keepaspectratio]{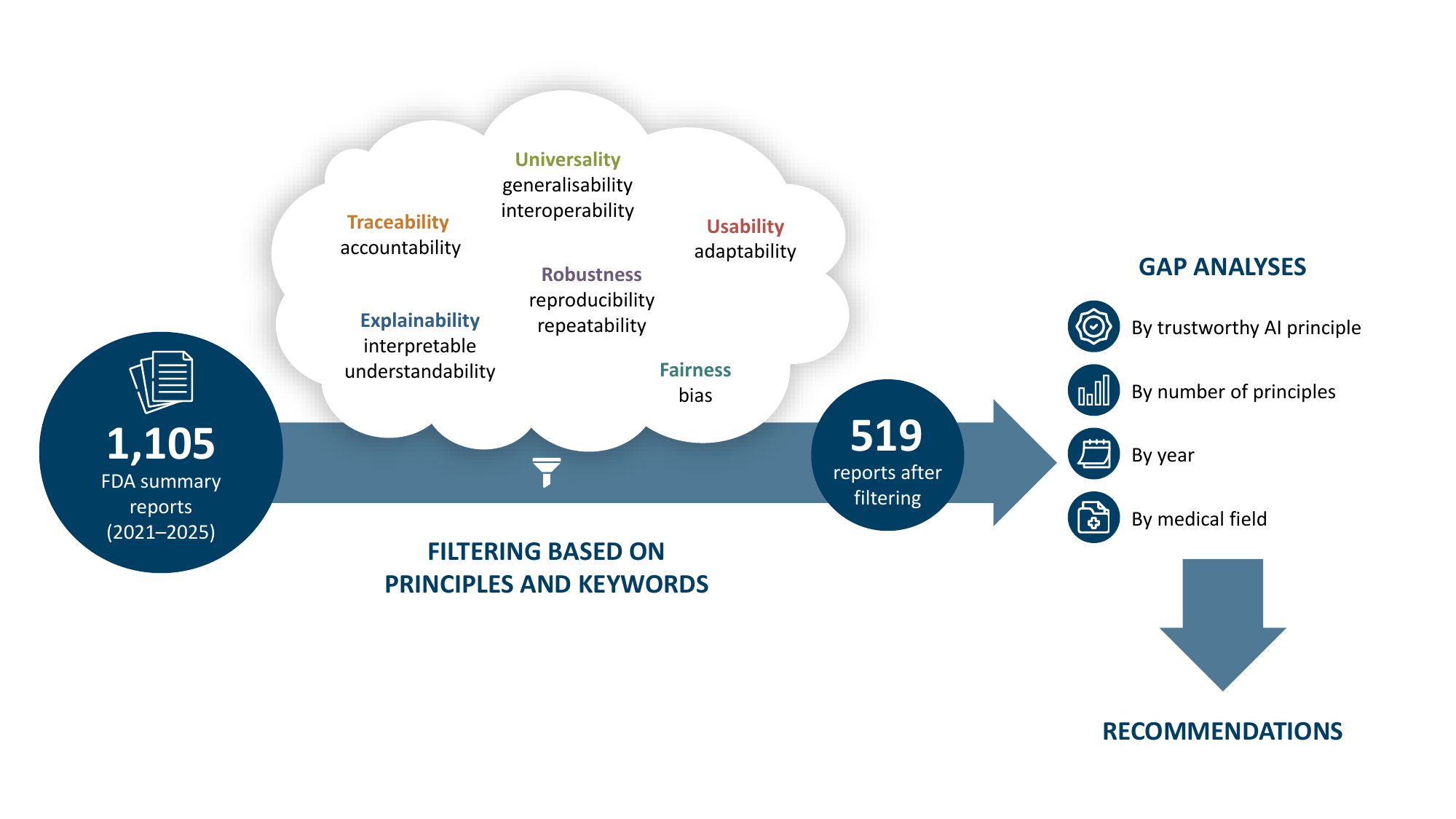}
    \caption{Study selection and analytical framework for assessing trustworthy AI reporting.}
    \label{Review process}
\end{figure}

Figure~\ref{Review process} summarizes the selection process of FDA AI/ML-enabled medical device summary reports from 2021 to 2025. A total of 1,105 reports were initially retrieved. Following automated keyword screening and manual validation, 586 reports were excluded because they did not contain evidence related to the predefined trustworthy AI criteria, resulting in 519 reports included in the final analysis.

\subsection{Evidence Across Trustworthy AI Principles}
\begin{figure}[H]
    \centering
    \includegraphics[width=\linewidth, keepaspectratio]{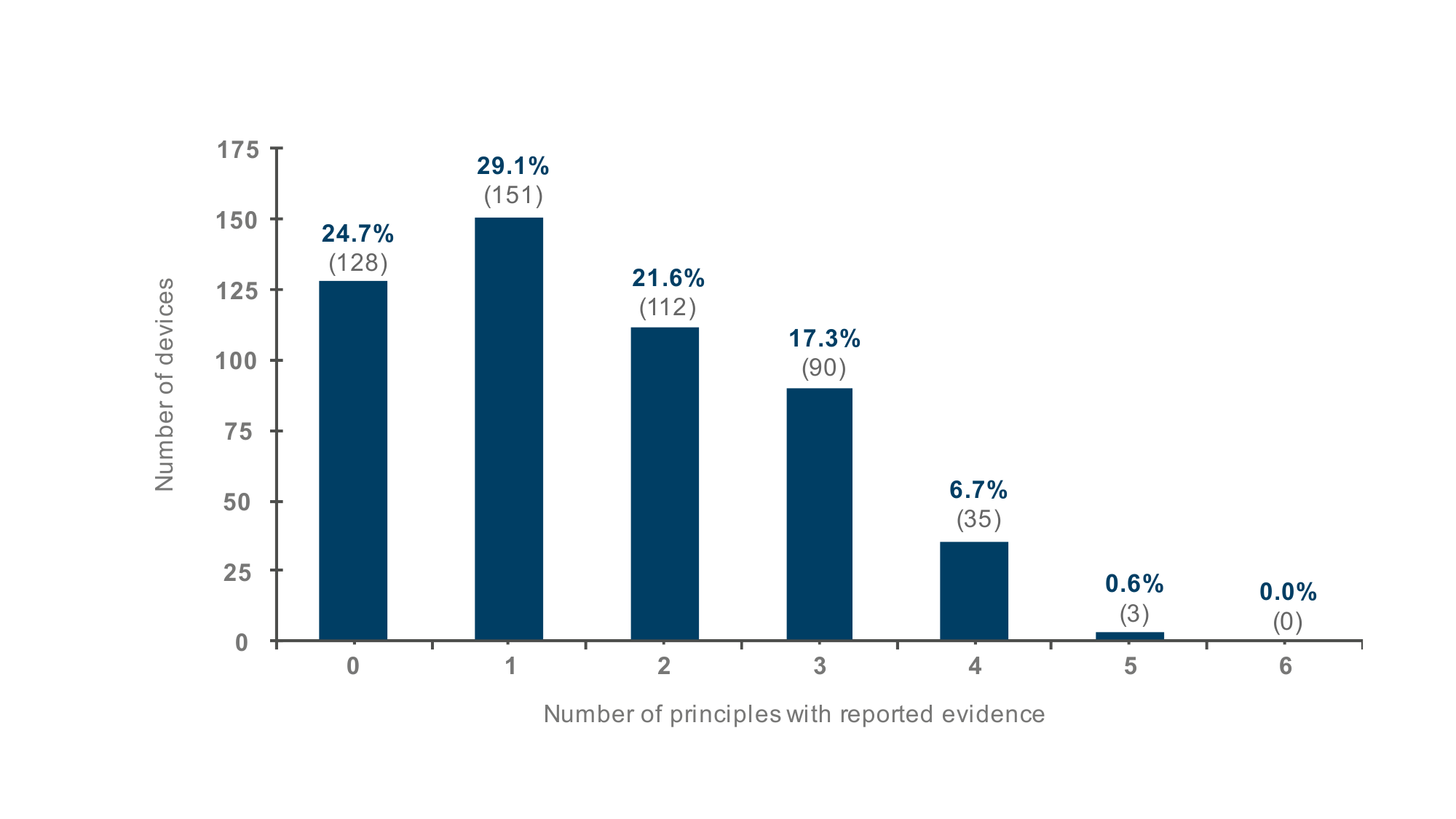}
    \caption{Distribution of FDA AI/ML summary reports by number of documented trustworthy AI principles.}
    \label{figure_3}
\end{figure}

Analysis of the 519 FDA AI/ML-enabled medical device summary reports revealed a limited level of comprehensive trustworthy AI reporting. As shown in Figure~\ref{figure_3}, nearly one quarter of reports (24.7\%, 128/519) did not provide evidence for any FUTURE-AI principle, while most reports provided evidence for only one (29.1\%, 151/519) or two principles (21.6\%, 112/519). Reporting across three principles was observed in only 17.3\% of reports (90/519), whereas evidence spanning four (6.7\%, 35/519) or five principles (0.6\%, 3/519) was rare. Notably, no report provided evidence covering all six FUTURE-AI principles. Overall, these findings highlight that comprehensive trustworthy AI reporting remains limited.

\begin{figure}[H]
    \centering
    \includegraphics[width=\linewidth, keepaspectratio]{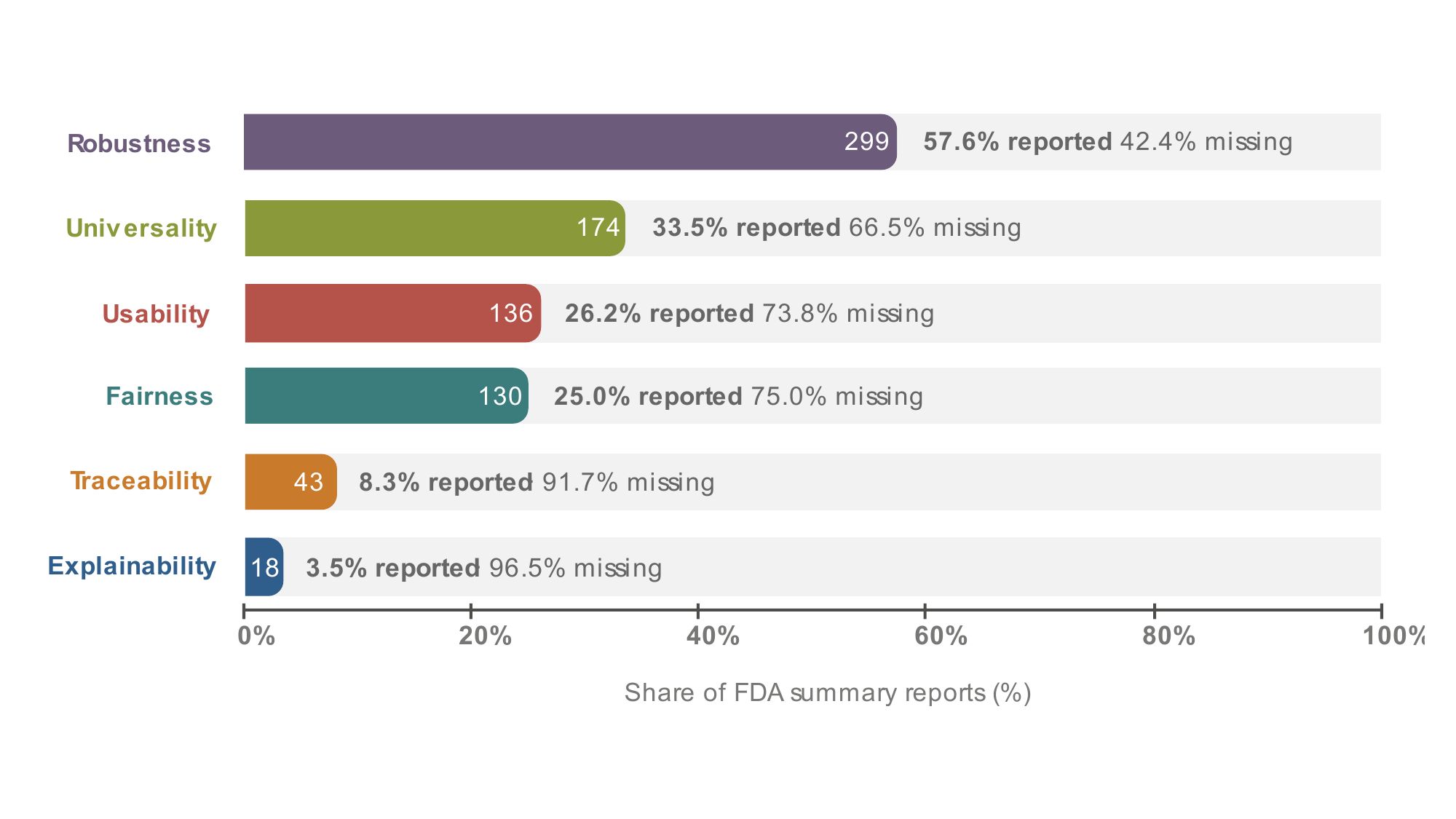}
    \caption{Distribution of documented evidence across the trustworthy AI principles..}
    \label{figure_1}
\end{figure}

\noindent
Figure~\ref{figure_1} shows the distribution of reporting across the six FUTURE-AI principles. Robustness was the most frequently documented principle, with evidence identified in 57.6\% of reports (299/519). In contrast, lower reporting rates were observed for the remaining principles, including Universality (33.5\%, 174/519), Usability (26.2\%, 136/519) and Fairness (25.0\%, 130/519). Traceability and Explainability represented the most pronounced reporting gaps, with evidence disclosed in only 8.3\% (43/519) and 3.5\% of reports (18/519), respectively. Overall, these findings reveal a strong imbalance in the reporting of trustworthy AI dimensions, with current FDA summary reports still prioritizing technical performance over broader ethical and human-centered dimensions of trustworthy AI.
\subsection{Trustworthy AI Reporting Across Clinical Domains}
\begin{figure}[H]
    \centering
    \includegraphics[width=\linewidth, keepaspectratio]{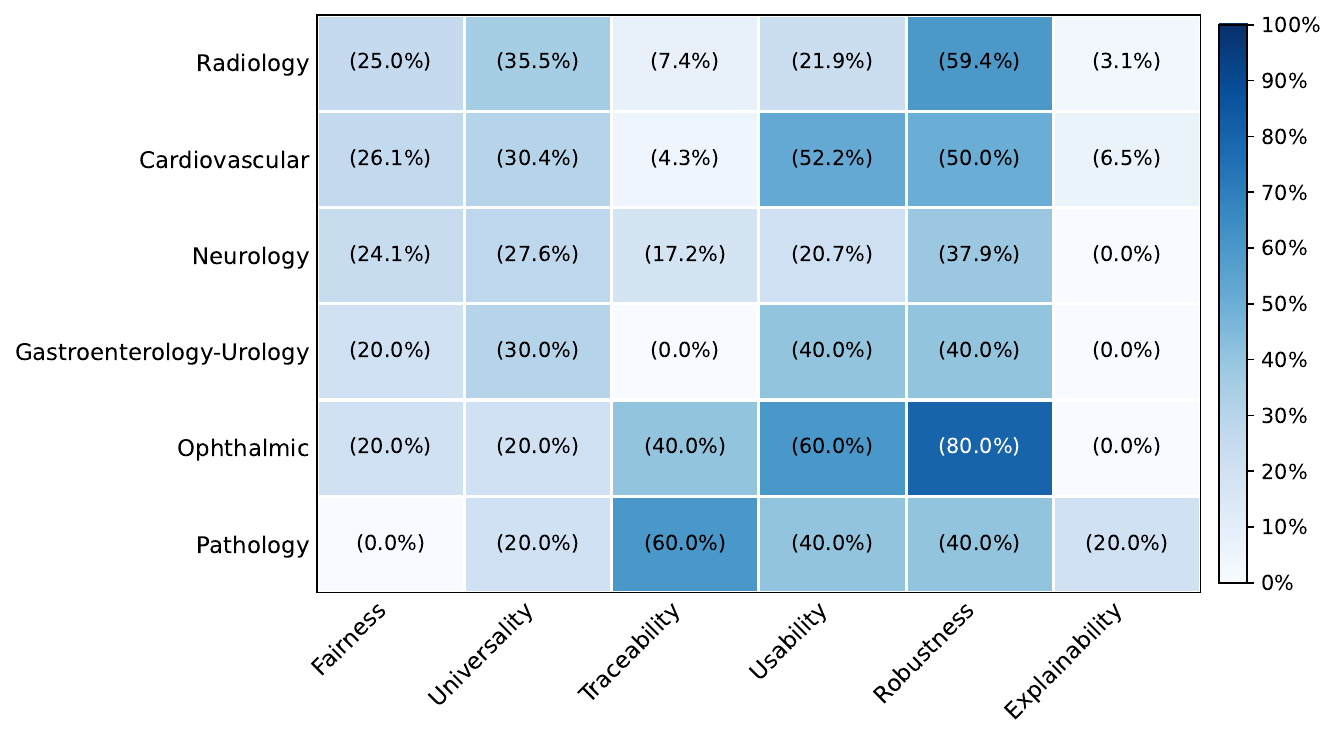}
    \caption{Distribution of documented evidence across trustworthy AI principles across prominent clinical domains.}
    \label{heatmap}
\end{figure}

Figure~\ref{heatmap} presents a heatmap summarising the distribution of reported evidence related to individual FUTURE-AI principles across six clinical domains. Robustness is the most consistently reported principle, reaching its highest concentration in Ophthalmic (80.0\%), Radiology (59.4\%), and Cardiovascular (50.0\%) applications. Usability is more frequently reported in Cardiovascular (52.2\%) and Ophthalmic (60.0\%) domains, where AI systems support interactive clinical workflows. In contrast, Traceability and Explainability are more prominent in Pathology (60.0\% and 20.0\%, respectively), reflecting its emphasis on auditability and case-based review. Fairness remains uniformly low across all domains, and Explainability is entirely absent in several specialties, indicating a cross-domain transparency gap. Overall, differences in trustworthy AI reporting appear driven by the nature of the principle and its clinical context, rather than by domain maturity or device volume alone.
\noindent
Furthermore, clinical domain was not significantly associated with reporting quality. In particular, compared with other medical specialties, Radiology reports showed no significant difference in the likelihood of high reporting transparency (OR 0.73, 95\% CI 0.46--1.15, p=0.17). These findings suggest that the "trustworthy AI" gap is not confined to specific clinical domains, but instead represents a broader systemic issue across the FDA AI/ML-enabled medical device landscape.
\subsection{Temporal Trends in Trustworthy AI Reporting}
\begin{figure}[H]
    \centering
    \includegraphics[width= \linewidth, keepaspectratio]{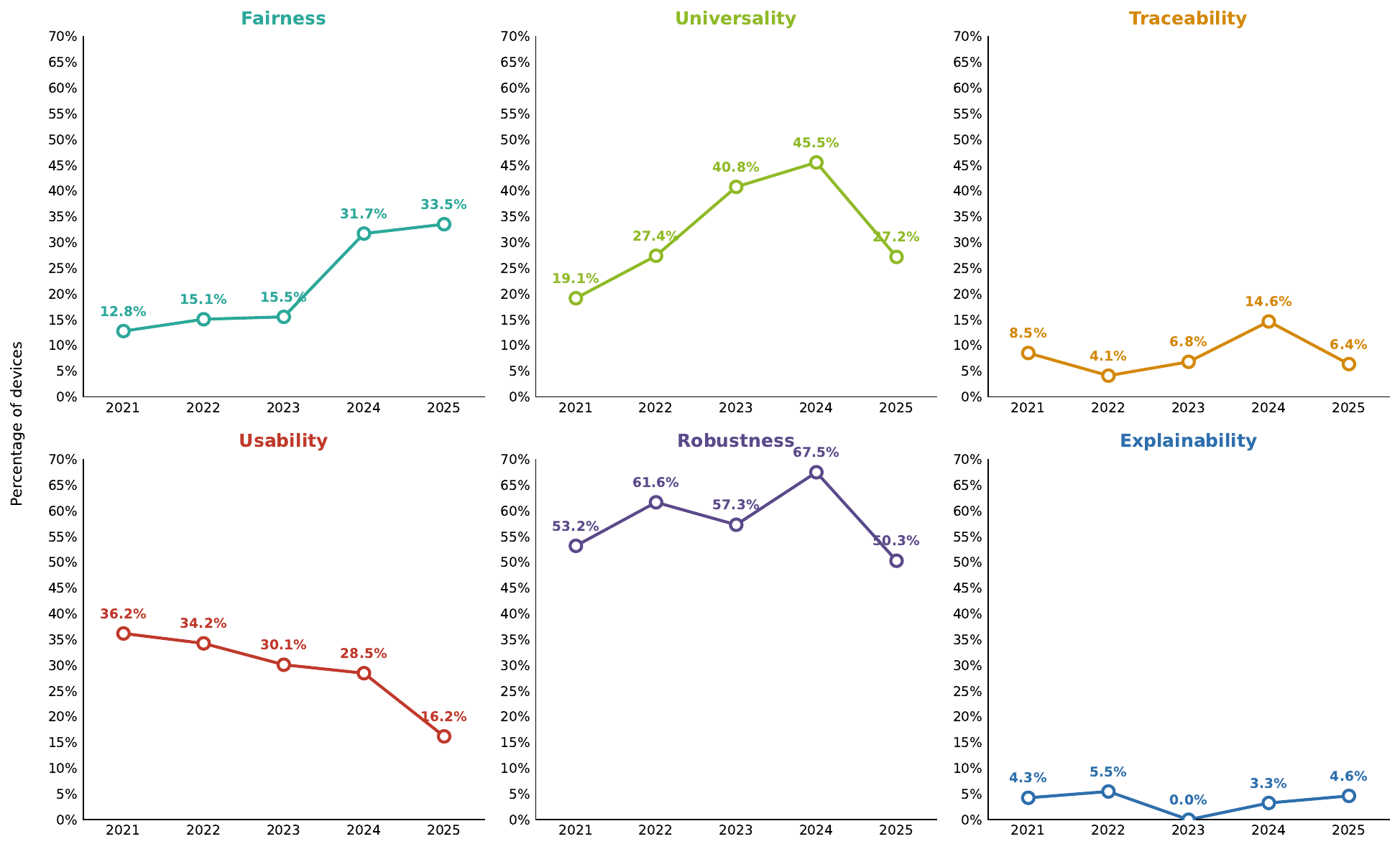}
    \caption{Temporal trends in reporting across trustworthy AI principles.}
    \label{TAI_group}
\end{figure}

Figure~\ref{TAI_group} presents the yearly proportion of FDA AI/ML summary reports containing documented evidence for each FUTURE-AI principle. Robustness remained the most consistently reported principle throughout the study period, ranging between 50.3\% and 67.5\%. Fairness showed a clear upward trajectory, rising from 12.8\% in 2021 to 33.5\% in 2025, while Universality peaked at 45.5\% in 2024 before declining sharply. Notably, Usability exhibited a steady decline from 36.2\% to 16.2\%, suggesting diminishing explicit documentation of human-factors considerations over time. Traceability remained persistently low and volatile, and Explainability was the most critically underreported principle across all years, never exceeding 5.5\% and dropping to 0.0\% in 2023. These divergent trajectories suggest that performance-oriented principles have become more embedded in regulatory workflows, while transparency-related dimensions remain structurally absent from the majority of submissions.
\noindent
A multivariable logistic regression using “High Transparency” (defined as reporting evidence for $\geq$ 3 trustworthy AI principles) as the dependent variable showed that year of FDA clearance was not significantly associated with higher reporting transparency (OR 1.02, 95\% CI 0.88--1.19, p=0.71). This result suggests that despite the increasing clearance of AI/ML-enabled devices over time, the comprehensiveness of trustworthy AI reporting has not meaningfully improved during the study period.
\subsection{Technical Evidence and Evaluation Strategies}
Beyond the frequency of reporting, the analysis also revealed substantial differences in how individual trustworthy AI principles were operationalized within FDA AI/ML public summary reports. Rather than evaluating the adequacy of the evidence itself, this section characterizes the types of methodological evidence typically disclosed for each principle and highlights recurring reporting limitations, even when principles are addressed.\\
Robustness and Universality were most commonly supported through retrospective validation studies using independent datasets, multi-site evaluations, subgroup analyses, and hardware compatibility testing. Robustness-related evidence primarily focused on repeatability, reproducibility, and test–retest reliability, whereas Universality emphasized generalizability across institutions, equipment, acquisition settings, or clinical environments. 
Fairness-related evidence was typically based on subgroup performance analyses across demographic or clinical populations, although explicit fairness metrics or bias quantification methods were rarely reported. Traceability evidence was generally documentation-oriented, relying on references to technical records, traceability matrices, verification and validation (V\&V) procedures, version control, or audit-related documentation. Usability reporting frequently focused on compliance with broader human factors and usability engineering standards, such as IEC 62366, although AI-specific evaluations of clinician interaction, trust, or workflow integration were less commonly described. Explainability represented the least reported dimension, with only five submissions explicitly using the term “explainability" and 18 devices disclosing explainability-related methods, most commonly through generic visualizations, saliency maps, or highlighted regions of interest.\\
Overall, our findings suggest that current reporting practices prioritize model performance, technical validation, and documentation, while standardized evaluation and reporting approaches for fairness, explainability, and other human-centered trustworthy AI dimensions remain limited.

\section{Discussion}
This study provides, to our knowledge, the first detailed assessment of trustworthy AI reporting within FDA AI/ML-enabled medical device summary reports. We used the multi-dimensional FUTURE-AI framework to systematically evaluate the reporting of key trustworthiness dimensions, including fairness, universality, traceability, usability, robustness, and explainability, which are critical for fostering public trust and supporting the safe adoption of AI systems in healthcare. Through the analysis of 519 FDA summary reports, we identified substantial reporting gaps across several trustworthy AI principles and revealed a marked disconnect between high-level trustworthy AI recommendations ~\cite{FDA_regu} and the level of evidence currently disclosed in publicly accessible regulatory documentation.
\subsection{Key Findings and Persistent Transparency Gaps}
Three main findings emerge from this analysis. First, trustworthy AI reporting remains sparse and fragmented across FDA AI/ML-enabled medical device summary reports. Nearly one quarter of reports did not provide evidence for any FUTURE-AI principle, while most documented evidence for only one or two principles. Furthermore, no report provided evidence covering all six FUTURE-AI dimensions. These findings suggest that publicly available regulatory documentation rarely supports a comprehensive assessment of trustworthiness, instead providing isolated evidence for selected aspects of AI system development and validation.\\
Second, reporting is heavily skewed toward technical performance-related dimensions of trustworthy AI. Robustness was by far the most frequently reported principle, whereas Explainability and Traceability were documented in fewer than 10\% of reports. This imbalance indicates that current reporting practices continue to prioritize evidence related to model performance, reproducibility, and validation while providing limited insight into how models generate predictions, how they can be audited, or how they are monitored throughout their lifecycle. Although substantial progress has been made within the research community to develop methods for explainability, fairness assessment, and human-centred evaluation, these advances appear to have only limited representation in publicly available FDA documentation.\\
Third, the identified transparency gap appears to be systemic rather than confined to specific clinical specialties or time periods. Reporting patterns were remarkably consistent across clinical domains, and neither year of clearance nor clinical specialty was significantly associated with higher reporting transparency. While some trustworthy AI dimensions, such as Fairness, showed gradual increases in reporting over time, similar progress was not observed across the remaining principles. This suggests that the increasing regulatory and policy focus on trustworthy AI has not yet translated into measurable changes in the depth or breadth of evidence disclosed in FDA public summaries.\\
It is important to emphasise that the identified reporting gaps also have broader ethical implications. Limited reporting of fairness, explainability, traceability, and usability may hinder the assessment of key ethical considerations such as equity, transparency, accountability, and human oversight, which are increasingly recognized as essential components of responsible AI in healthcare. Consequently, insufficient evidence regarding trustworthiness may not only limit the technical evaluation of AI systems, but also impede independent assessment of whether they align with emerging ethical and societal expectations for healthcare AI.\\
A few limitations should be considered when interpreting these findings. First, our analysis was restricted to publicly available FDA summary reports and did not include confidential regulatory submissions or internal manufacturer documentation. Consequently, the absence of evidence for a given trustworthy AI principle should not be interpreted as definitive evidence of non-compliance. Second, although a consensus-based review process was implemented, the binary assessment framework may not fully capture differences in the depth or quality of reported evidence. Nevertheless, from the perspective of transparency, these limitations do not alter the central finding that publicly available documentation often provides insufficient information to independently assess the trustworthiness of approved AI systems. 
\subsection{Differential Adoption of Trustworthy AI Principles}
Although the principles of trustworthy AI as defined by the FUTURE-AI framework are increasingly recognized as complementary and equally important dimensions of trustworthy AI in healthcare, our findings suggest that their adoption within regulatory reporting remains highly uneven. Examining each principle individually in our study provides additional insight into which dimensions of trustworthy AI have become more embedded within current regulatory practice and which continue to face barriers to implementation, evaluation, or transparent reporting.

\begin{itemize}
    \item \textbf{Robustness:} The high reporting of robustness is unsurprising given its close alignment with traditional regulatory expectations around performance validation, reproducibility, reliability, and testing. Robustness evidence is also comparatively straightforward to quantify using established statistical and validation methodologies.

    \item \textbf{Universality:} The relatively high degree of universality reflects the recognition that AI systems must generalize beyond their development environments. However, most evidence focused on external validation or multi-site testing, with limited discussion of broader interoperability or deployment across heterogeneous healthcare settings, especially in resource-constrained settings.

    \item \textbf{Fairness:} The gradual increase in fairness reporting is encouraging and may reflect growing regulatory attention to bias and equity. Nevertheless, fairness assessments were often limited to subgroup performance comparisons, with little evidence of more advanced bias detection, mitigation, or monitoring strategies.

    \item \textbf{Traceability:} The low reporting of traceability is concerning because traceability forms the foundation of regulatory auditability and lifecycle oversight. The absence of detailed information regarding model provenance, monitoring, and update procedures may limit the ability of healthcare organizations to understand how AI systems evolve over time.

    \item \textbf{Usability:} Although usability was reported more frequently than explainability or traceability, much of the evidence focused on conventional human-factors engineering requirements rather than AI-specific evaluations of clinician trust, workflow integration, or decision support effectiveness. This suggests that human-centered deployment remains incompletely assessed.

    \item \textbf{Explainability}: Explainability represents the most striking reporting gap. Despite substantial advances in explainable AI research, there remains challenges on how explanations should be validated, how their clinical value should be demonstrated, and how such evidence should be reported within regulatory submissions. As a result, explainability appears to be one of the least operationalized dimensions of trustworthy AI in practice.

\end{itemize}
   
\subsection{Healthcare Implications of Limited Trustworthy AI Reporting}
Although this study focuses on public regulatory documentation rather than clinical outcomes, the identified reporting gaps have important implications for the adoption and use of AI in healthcare. First, limited transparency may reduce clinician trust and willingness to integrate AI systems into clinical decision-making when insufficient evidence is available regarding how models were developed, validated, and monitored. Trust is particularly important in high-stakes clinical settings, where AI recommendations may directly influence diagnosis, treatment planning, or patient management.\\
Second, incomplete reporting increases the risk of inappropriate use and potential patient harm. Without transparent evidence regarding potential biases, failure modes, usability constraints, explainability mechanisms, and monitoring procedures, AI systems may be deployed in settings or populations for which they were not adequately evaluated. This may limit the ability of clinicians and healthcare organizations to identify situations in which model performance could deteriorate, increasing the risk of unreliable outputs, inappropriate decisions, and clinical errors.\\
Finally, limited reporting of trustworthy AI evidence complicates the objective comparison of competing AI solutions. Healthcare organizations, procurement teams, and clinical leaders increasingly require evidence not only of performance but also of reliability, fairness, and usability when selecting AI technologies. The absence of standardized reporting of AI trustworthiness therefore limits informed decision-making and may hinder the responsible adoption of AI-enabled medical devices.

\subsection{Policy and Regulatory Implications}
As summarized in Table~\ref{tab:regulatory_mapping}, there is already strong alignment between the trustworthy AI principles as defined by the FUTURE-AI international guideline and emerging regulatory and policy frameworks, such as the FDA’s Good Machine Learning Practice (GMLP), the EU AI Act, and US Executive Order 14110. Despite this convergence, our audit reveals a substantial gap between these high-level recommendations and the trustworthy AI evidence currently disclosed in public regulatory documentation. These findings highlight the need to move beyond voluntary, heterogeneous reporting practices toward more systematic, standardized, and audit-ready disclosure of trustworthy AI evidence. Based on the gaps identified in this study, we propose the following recommendations:\\
\textbf{Recommendation 1: Comprehensive Reporting Across All Trustworthy AI Principles}
Current reporting practices appear to disproportionately emphasize technical performance, robustness and external testing, while providing limited evidence regarding other key dimensions of trustworthy AI, including fairness, explainability, traceability, and usability. Future regulatory summaries should therefore encourage balanced reporting across all trustworthy AI dimensions rather than selectively documenting those that are easiest to evaluate or most closely aligned with traditional performance assessment. Such an approach would provide clinicians, healthcare organizations, and regulators with a more complete understanding of the strengths, limitations, risks, and deployment readiness of AI-enabled medical devices.\\
\textbf{Recommendation 2: Standardized Trustworthy AI Reporting Templates and Terminology}
The substantial heterogeneity observed across FDA summary reports highlights the need for more standardized reporting practices. We propose that future regulatory summaries include a dedicated Trustworthy AI section structured around clearly defined dimensions of trustworthy AI, such as the six FUTURE-AI principles. Such a framework could be accompanied by standardized terminology, definitions, and minimum reporting requirements to improve consistency, comparability, and interpretability across devices and manufacturers. Similar to reporting guidelines such as CONSORT-AI, SPIRIT-AI, and TRIPOD-AI, a structured trustworthy AI reporting template could facilitate more transparent communication of evidence while reducing ambiguity in the interpretation of reported claims.\\
\textbf{Recommendation 3: Greater Emphasis on Human-Centred Evaluation}
Our findings suggest that usability-related evidence remains limited and is often restricted to compliance with general human-factors standards rather than AI-specific evaluation of human-AI interaction. Future regulatory reporting should place greater emphasis on human-centred evidence, including clinician engagement, workflow integration, user understanding of model outputs, and the impact of AI systems on clinical decision-making. This is particularly relevant for explainability and usability, where the ultimate value of an AI system depends not only on algorithmic performance but also on how effectively clinicians can understand, trust, and appropriately use its outputs in real-world practice.\\
\textbf{Recommendation 4: Assessing Trustworthiness Throughout the AI Lifecycle}
Unlike traditional medical devices, AI systems are dynamic technologies whose trustworthiness may evolve over time as clinical environments, patient populations, data distributions, and usage patterns change. A model that demonstrates acceptable levels of fairness, robustness, or generalizability at the time of regulatory approval may not necessarily maintain these properties after deployment. Trustworthiness should therefore not be viewed as a static requirement established at regulatory approval, but as a dynamic characteristic that must be continuously anticipated, monitored, and maintained throughout the AI lifecycle. Future regulatory summaries should therefore include pre-deployment assessments of temporal stability and vulnerability to changing conditions, including evaluations across evolving datasets, populations, devices, and care settings, while clearly documenting potential sources of performance degradation. Manufacturers should also describe device-specific post-deployment monitoring strategies, including for tracking model performance, detecting data and population drift, monitoring subgroup fairness, or performing quality control of AI inputs and outputs. They should also report mitigation measures that can be activated when degradations in trustworthiness are detected, such as recalibration procedures, model updating policies, bias mitigation approaches, human oversight mechanisms, and governance frameworks for continuous learning systems.

\begin{table}[H]
\centering
\caption{Mapping of FUTURE-AI principles to major international regulatory frameworks and guidelines. FAI: FUTURE-AI; US: United State; UK: United Kingdom; EU: European Union.}
\label{tab:regulatory_mapping}
\small
\begin{tabular}{|>{\raggedright\arraybackslash}p{2cm} |>{\raggedright\arraybackslash}p{4.1cm} |>{\raggedright\arraybackslash}p{4.1cm} | >{\raggedright\arraybackslash}p{3.5cm}|}
\hline
\textbf{FAI Principles} & \textbf{US/UK/Canada GMLP (2021)} & \textbf{EU AI Act (2024)} & \textbf{US Exec. Order 14110 (2023)} \\ \hline
\textbf{Fairness} & \textbf{Principle 3:} Data sets are representative of the intended patient population to manage bias. & \textbf{Art. 10:} Data governance and bias monitoring requirements. & \textbf{Sec. 7:} Ensuring equity and preventing algorithmic discrimination. \\ \hline
\textbf{Universality} & \textbf{Principle 3 and 5:} Selected Reference Datasets Are Based Upon Best Available Methods. & \textbf{Art. 13:} Transparency and provision of information to users. & \textbf{Sec. 4:} Standards for AI safety and interoperability. \\ \hline
\textbf{Traceability} & \textbf{Principle 9:} Users Are Provided Clear, Essential Information. & \textbf{Art. 12:} Automatic recording of events (logging) and technical documentation. & \textbf{Sec. 4.1:} Requirements for auditable and transparent AI. \\ \hline
\textbf{Usability} & \textbf{Principle 2:} Good Software Engineering and Security Practices Are Implemented. & \textbf{Art. 14:} Human oversight to prevent or minimize risks (Human-in-the-loop). & \textbf{Sec. 8:} Protecting consumers and healthcare patients. \\ \hline
\textbf{Robustness} & \textbf{Principle 4:} Training and test data sets are independent; testing in relevant conditions. & \textbf{Art. 15:} Accuracy, robustness, and cybersecurity requirements. & \textbf{Sec. 4.1:} Developing rigorous standards for red-team testing. \\ \hline
\textbf{Explainability} & \textbf{Principle 7:} Focus on the performance of the Human-AI team and human interpretability. & \textbf{Art. 13:} Interpretability to allow users to understand and use outputs. & \textbf{Sec. 4.1:} Promoting transparency in AI development and use. \\ \hline
\end{tabular}
\end{table}

\section{Conclusion}
This study provides the first systematic assessment of trustworthy AI evidence in FDA AI/ML-enabled medical device summary reports. Using the FUTURE-AI framework, we evaluated 519 reports and identified substantial reporting gaps across all dimensions of trustworthy AI, with limited improvements observed over time and no meaningful differences across clinical specialties. These findings reveal a persistent disconnect between emerging trustworthy AI expectations and the evidence currently disclosed in public regulatory documentation.\\
As AI becomes increasingly integrated into healthcare, regulatory approval alone should not be considered a proxy for trustworthiness or guaranteed safety. We advocate for standardized, audit-ready reporting of trustworthy AI evidence across all dimensions and throughout the AI lifecycle. Such transparency would enable clinicians, healthcare organizations, regulators, and patients to independently assess the robustness, fairness, usability, explainability, and real-world deployability of each AI-enabled medical device.

\section*{Author Contributions}
Ahmed M Salih: conceptualization, data curation, formal analysis, methodology, project administration, validation, writing – original draft; Oliver Díaz, Alejandro Guzman, Noah Marquez Vara, Fotios Avgoustidis, Rituraj Singh and Saman Barakat: data curation, formal analysis, methodology, validation, writing – review and editing. Zahra Raisi-Estabragh and Karim Lekadir: supervision, validation, writing – review and editing.
\section*{Acknowledgments}
This work is supported by the European Union’s Horizon Europe research and innovation programme under grant agreement No 101233553 (COMPASS-AI). AMS acknowledges the support of Leicester City Football Club (LCFC) and from British Heart Foundation (RE/24/130031). OD is partially supported by the project AIMED (PID2023-146786OB-I00) from the Ministry of Science and Innovation of Spain. RS is fully supported by the European Union’s Horizon Europe research and innovation programme under the Marie Skłodowska-Curie grant agreement No 101211445 (BRIDGE project). During the preparation of this work the author(s) used the free plan of ChatGPT-5 in order to assist with language clarity, flow, and grammatical accuracy. After using this tool/service, the author(s) reviewed and edited the content as needed and take(s) full responsibility for the content of the published article.
\section*{Competing interests}
All authors declare no conflicts of interest.
\section*{Data availability}
All the data used in the tutorial are publicly available and can be accessed through the FDA approved AI-based medical devices portal.
\section*{Code availability}
All codes generated in the manuscript can be available upon request from the authors.

\printbibliography

\newpage
\section*{Supplementary Tables}

\renewcommand{\tablename}{Supp Table}
\setcounter{table}{0}
\begin{table}[H]
\small
\caption{Matching FUTURE-AI principles evaluation} \label{TAI_ev}
\begin{tabular}{|p{7cm}|p{8cm}|}
\hline
{\color[HTML]{1F1F1F} \textbf{Cateogry}}                                    & {\color[HTML]{1F1F1F} \textbf{Match}}                                                                                                  \\ \hline
{\color[HTML]{1F1F1F} Subgroup   analyses and multi-site diversity}         & {\color[HTML]{1F1F1F} Different sites,   diverse cohorts, multiple devices, age/gender diversity}                                      \\ \hline
{\color[HTML]{1F1F1F} Human-AI   comparison and manual validation}          & {\color[HTML]{1F1F1F} Comparison with   manual segmentation, radiologist benchmarks, expert panels}                                    \\ \hline
{\color[HTML]{1F1F1F} Repeatability   and test-retest reliability}          & {\color[HTML]{1F1F1F} Test-retest, multiple   runs, intraclass correlation (ICC), stability checks}                                    \\ \hline
{\color[HTML]{1F1F1F} Technical   documentation and traceability}           & {\color[HTML]{1F1F1F} Traceability matrix,   V\&V analysis, code review, reference to specific papers/docs}                            \\ \hline
{\color[HTML]{1F1F1F} Bias   mitigation and fairness checks}                & {\color[HTML]{1F1F1F} Randomization of   order, removing confounding variables, unbiased data checks}                                  \\ \hline
{\color[HTML]{1F1F1F} Model   explainability and visualization}             & {\color[HTML]{1F1F1F} SHAP, heatmaps,   Regions of Interest (ROI), confidence scores, probability maps}                                \\ \hline
{\color[HTML]{1F1F1F} Performance   metrics and validation datasets}        & {\color[HTML]{1F1F1F} Specific accuracy   metrics, independent test sets, data drift monitoring}                                       \\ \hline
{\color[HTML]{1F1F1F} System   interoperability and hardware compatibility} & {\color[HTML]{1F1F1F} DICOM standards, OS   requirements (CentOS/Windows), cross-platform testing}                                     \\ \hline
{\color[HTML]{1F1F1F} Usability and   user experience evaluation}           & {\color[HTML]{1F1F1F} Summative usability   tests, user feedback studies, simulated use}                                               \\ \hline
{\color[HTML]{1F1F1F} Insufficient   information / Not provided}            & {\color[HTML]{1F1F1F} "No details   mentioned", "Nothing mentioned", "No information"}                                                 \\ \hline
\end{tabular}
\end{table}

\end{document}